\documentclass[12pt] {article}
\usepackage{psfrag}
\usepackage{graphicx}
\usepackage{latexsym,amsfonts}
\usepackage{amssymb}
\begin{document}
\setcounter{page}{1}
\def\theequation{\arabic{section}.\arabic{equation}}
\def\theequation{\thesection.\arabic{equation}}
\setcounter{section}{0}

\title{On radiative $np \to 1s + \gamma$ transitions, \\induced by
  strong low--energy interactions,\\ in kaonic atoms}

\author{A. N. Ivanov\,\thanks{E--mail: ivanov@kph.tuwien.ac.at, Tel.:
+43--1--58801--14261, Fax: +43--1--58801--14299}~\thanks{Permanent
Address: State Polytechnic University, Department of Nuclear Physics,
195251 St. Petersburg, Russian Federation}\,,
M. Cargnelli\,\thanks{E--mail: michael.cargnelli@oeaw.ac.at}\,,
M. Faber\,\thanks{E--mail: faber@kph.tuwien.ac.at, Tel.:
+43--1--58801--14261, Fax: +43--1--58801--14299}\,, H.
Fuhrmann\,\thanks{E--mail: hermann.fuhrmann@oeaw.ac.at}\,,\\
V. A. Ivanova\,\thanks{E--mail: viola@kph.tuwien.ac.at, State
Polytechnic University, Department of Nuclear Physics, 195251
St. Petersburg, Russian Federation}\,, J. Marton\,\thanks{E--mail:
johann.marton@oeaw.ac.at}\,, N. I. Troitskaya\,\thanks{State
Polytechnic University, Department of Nuclear Physics, 195251
St. Petersburg, Russian Federation}~~, J. Zmeskal\,\thanks{E--mail:
johann.zmeskal@oeaw.ac.at}} 
\date{\today}

\maketitle

\begin{center}
{\it Atominstitut der \"Osterreichischen Universit\"aten,
Arbeitsbereich Kernphysik und Nukleare Astrophysik, Technische
Universit\"at Wien, \\ Wiedner Hauptstr. 8-10, A-1040 Wien,
\"Osterreich \\ and\\ Stefan Meyer Institut f\"ur subatomare Physik 
der \"Osterreichische Akademie der Wissenschaften,\\
Boltzmanngasse 3, A-1090, Wien, \"Osterreich} 
\end{center}

\begin{center}
\begin{abstract}
  We calculate the rates of the radiative transitions $np \to 1s +
  \gamma$ in kaonic hydrogen and kaonic deuterium, induced by strong
  low--energy interactions and enhanced by Coulomb interactions.  The
  obtained results should be taken into account for the theoretical
  analysis of the experimental data on the $X$--ray spectra and yields
  in kaonic atoms.\\
  \vspace{0.1in}
  \end{abstract}

\end{center}

\newpage

\section{Introduction}
\setcounter{equation}{0}

The $X$--ray spectra and yields \cite{XR1}---\cite{XR7}, produced by
the atomic transitions $np \to 1s$ in kaonic hydrogen, where $n$ is
the {\it principle quantum number} of the energy levels, are the main
experimental tool for the measurement of the energy level displacement
of the ground state of kaonic hydrogen, caused by strong low--energy
$\bar{K}N$ interactions \cite{DEAR1}.  It is known
\cite{XR4}--\cite{XR7} that the $X$--ray yields related to the
$K_{\alpha}$, $K_{\beta}$ and $K_{\gamma}$ lines of kaonic hydrogen
are very sensitive to the value of $\Gamma_{2p}$, the rate of hadronic
decays of kaonic hydrogen from the $2p$ state. Usually $\Gamma_{2p}$
is used as an input parameter in the theories of the atomic cascades
\cite{XR4}--\cite{XR7}.

Recently \cite{IV5} we have calculated the rate $\Gamma_{np}$ of
hadronic decays of kaonic hydrogen from the $np$  state.  For
the $2p$ state we have obtained: $\Gamma_{2p} = 2\,{\rm meV} =
3.0\times 10^{12}\,{\rm sec^{-1}}$.  This agrees well with the
assumption by Koike, Harada, and Akaishi \cite{XR4}. In order to
reconcile the experimental data on the $K_{\alpha}$, $K_{\beta}$ and
$K_{\gamma}$ lines with the theoretical analysis they assumed that
$\Gamma_{2p} > 1\,{\rm meV} = 1.5\times 10^{12}\,{\rm sec^{-1}}$.

In this paper we continue the analysis of the influence of strong
low--energy interactions on the transitions from the  $np$
state in kaonic hydrogen, which we have started in \cite{IV5}. We
investigate the radiative transitions $np \to 1s + \gamma$, induced by
strong low--energy interactions and enhanced by the Coulomb
interaction of the $K^-p$ pair, in kaonic hydrogen and extend the
results on kaonic deuterium.

The paper is organized as follows. In Section 2 we calculate the rates
of the radiative transitions $np \to 1s + \gamma$ in kaonic hydrogen,
induced by strong low--energy interactions and enhanced by the Coulomb
interaction. The Coulomb interaction is taken into account in the form
of the explicit non--relativistic Coulomb wave functions of the
relative motion of the $K^-p$ pairs and the explicit non--relativistic
Coulomb Green functions for the calculation of the amplitude of the
kaon--proton {\it Bremsstrahlung} $K^-+ p \to K^- + p + \gamma$,
defining the rate of the radiative atomic transition $(K^-p)_{np} \to
(K^-p)_{1s} + \gamma$ in our approach. In Section 3 we modify the
rates of the transitions $np \to 1s + \gamma$, calculated in Section 2
for kaonic hydrogen, for kaonic deuterium.  In the Conclusion we
discuss the obtained results.

\section{$np \to 1s + \gamma$ transitions in kaonic
  hydrogen}
\setcounter{equation}{0}

The rate of the radiative transition $(K^-p)_{np} \to (K^-p)_{1s} +
\gamma$ can be defined by \cite{IV3}
\begin{eqnarray}\label{label2.1}
\hspace{-0.3in} \Gamma((K^-p)_{np} \to (K^-p)_{1s}\, \gamma) = 
\frac{1}{8\pi}\,\frac{\omega}{(m_K + m_N)^2}\,\overline{|M((K^-p)_{np}
\to (K^-p)_{1s}\,\gamma)|^2},
\end{eqnarray}
where $\omega = E_{np} - E_{1s} = \alpha^2 \mu (n^2 - 1)/2n^2$ is the
photon energy, $\alpha = 1/137.036$ is the fine--structure constant,
$\mu = m_K m_N/ (m_K + m_N) = 324\,{\rm MeV}$ is the reduced mass of
the $K^-p$ pair calculated for $m_K = 494\,{\rm MeV}$ and $m_N =
940\,{\rm MeV}$. The amplitude $M((K^-p)_{np} \to (K^-p)_{1s}\,
\gamma)$ is given by \cite{IV3}
\begin{eqnarray}\label{label2.2}
  M((K^-p)_{np} \to (K^-p)_{1s}\,\gamma) =
   \frac{1}{2\mu}
  \int \frac{d^3q}{(2\pi)^3} 
  \frac{d^3k}{(2\pi)^3}
  \Phi^*_{100}(\vec{k})\Phi_{n1m}(\vec{q}\,)M(K^- p \to K^- p\, \gamma),
\end{eqnarray}
where $\Phi_{100}(\vec{k}\,) = \Phi_{1s}(k)$ and $\Phi_{n1m}(\vec{q})
= -\,i\,\sqrt{4\pi}\,\Phi_{np}(q)\,Y_{1m}(\vartheta_{\vec{q}},
\varphi_{\vec{q}})$ are the wave functions of kaonic hydrogen in the
ground $1s$ and the  $np$ state in the momentum representation
\cite{BS57}; $Y_{1m}(\vartheta_{\vec{q}}, \varphi_{\vec{q}})$ is a
spherical harmonic and $\Phi_{1s}(k)$ and $\Phi_{np}(q)$ are radial
wave functions of kaonic hydrogen in the momentum representation
\cite{BS57} (see also \cite{IV5,IV2}). Then, $M(K^- p \to K^-
p\,\gamma)$ is the amplitude of the kaon--proton {\it Bremsstrahlung}
 $K^- + p \to K^- + p + \gamma$.

 The amplitude $M(K^- p \to K^- p\,\gamma)$ is defined by the Feynman
 diagrams depicted in Fig.1. For the calculation of this amplitude we
 use the following effective Lagrangian, describing strong low--energy
 and electromagnetic interactions of the $K^-p$ pairs:
\begin{eqnarray}\label{label2.3}
  {\cal L}_{\rm eff}(x) &=& \partial_{\mu}K^{-\dagger}(x)\partial^{\mu}K^-(x) 
  - m^2_KK^{-\dagger}(x) K^-(x) + \bar{p}(x)(i\gamma^{\mu}\partial_{\mu} - m_N)p(x)
  \nonumber\\
  && + ie\,(K^{-\dagger}(x)\partial_{\mu}K^-(x) - 
  \partial_{\mu}K^{-\dagger}(x)\,K^-(x))\,A^{\mu}(x) - 
  e\,\bar{p}(x)\gamma_{\mu}p(x)\,A^{\mu}(x)\nonumber\\
  && + 4\pi\Big(1 + \frac{m_K}{m_N}\Big)\,a^{K^-p}_0\,
  \bar{p}(x)p(x)K^{-\dagger}(x)K^-(x) + \ldots, 
\end{eqnarray}
where $e$ is the electric charge of the proton such as $e^2 = 4\pi
\alpha$, $a^{K^-p}_0$ is the S--wave scattering length of $K^-p$
scattering \cite{IV3,IV4}, $A^{\mu}(x)$ is the vector potential of the
quantized electromagnetic field. The ellipses denote interactions of
order of $O(e^2)$, which we omit. In the effective Lagrangian
Eq.(\ref{label2.3}) strong low--energy interactions of the $K^-p$ pair
are described by ${\cal L}_{KKpp}(x) = 4\pi (1 + m_K/m_N)\, a^{K^-p}_0\,
\bar{p}(x)p(x)K^{-\dagger}(x)K^-(x)$.

\begin{figure}
  \centering \psfrag{K}{$K^-$}
  \psfrag{p}{$p$}\psfrag{g}{$\gamma$}
  \includegraphics[height=0.30\textheight]{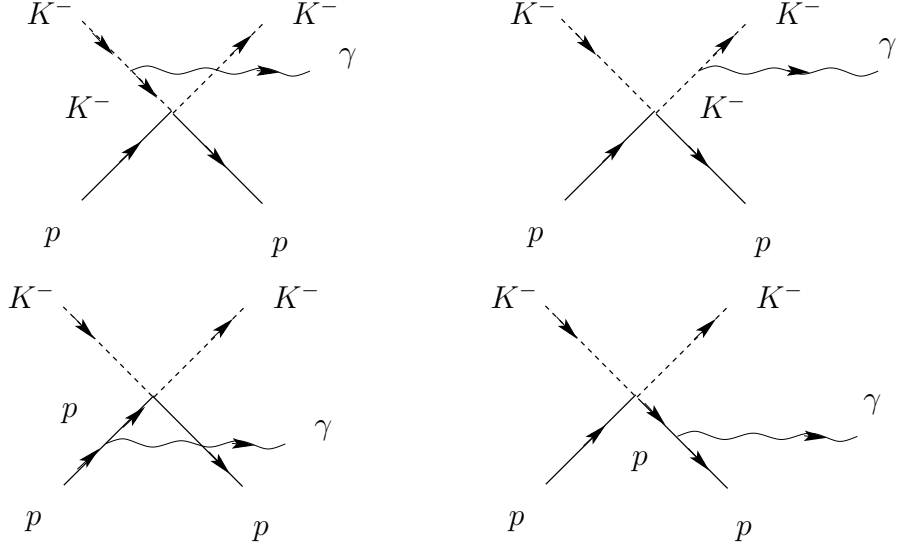}
\caption{Feynman diagrams for the radiative transitions $(K^-p)_{np} \to 
(K^-p)_{1s} + \gamma$ caused by strong low--energy interactions} 
\end{figure}

In the non--relativistic limit the amplitude of the kaon--proton {\it
  Bremsstrahlung} $K^- + p \to K^- + p + \gamma$ in Fig.1 reads
\begin{eqnarray}\label{label2.4}
  \hspace{-0.3in}M(K^- p \to K^- p \, \gamma) = 8\pi\,(m_K + m_N)\,a^{K^-p}_0\,
  \frac{i e}{2\mu \omega}\,
  \vec{e}^{\;*}(\vec{p},\lambda)\cdot( \vec{k} +  \vec{q}\,).
\end{eqnarray}
For the calculation of the amplitude of the atomic transition
$(K^-p)_{np} \to (K^-p)_{1s} + \gamma$ the amplitude of the
kaon--proton {\it Bremsstrahlung}  Eq.(\ref{label2.4}) should
be weighted with the wave functions of kaonic hydrogen in the ground
and excited $np$ states Eq.(\ref{label2.2}). Since the wave function
of the ground state is spherical symmetric, the integration over
$\vec{k}$ leads to the vanishing of the term proportional to
$\vec{k}\cdot \vec{e}^{\;*}(\vec{p},\lambda)$.  Therefore, below we
omit it.

The appearance of the photon energy $\omega$ in the denominator of the
amplitude of the kaon--proton {\it Bremsstrahlung} $K^- +p \to K^- + p
+ \gamma$ is due to the fact that the virtual $K^-$--meson and the
proton are practically on--mass shell.

Our approach to the calculation of the amplitude of the kaon--proton
{\it Bremsstrahlung} $K^- + p \to K^- + p + \gamma$ is similar to that
which has been used in \cite{IV4} for the derivation of the
Ericson--Weise formula for the S--wave scattering amplitude of $K^-d$
scattering \cite{TE88} and runs parallel the non--relativistic
Effective Field Theory (the EFT) approach, based on Chiral
Perturbation Theory (ChPT) by Gasser and Leutwyler \cite{JG83}, which
has been applied by Mei\ss ner {\it et al.}  \cite{UM04} to the
systematic calculation of QCD isospin--breaking and electromagnetic
corrections to the energy level displacement of the $ns$ state of
kaonic hydrogen. In \cite{IV5} we have analysed the quantitative
agreement of the results on the energy level displacement of the $ns$
state of kaonic hydrogen obtained by Mei\ss ner' s group \cite{UM04},
our group \cite{IV3,IV4} and the experimental data by the DEAR
Collaboration \cite{DEAR1}.

Due to the Coulomb interaction of the $K^-p$ pair in initial and final
states \cite{TE77}--\cite{LL65} the amplitude of the kaon--proton {\it
  Bremsstrahlung} Eq.(\ref{label2.4}) changes as follows
\begin{eqnarray}\label{label2.5}
  &&M(K^-  p \to K^-  p \,\gamma) = 8\pi\,(m_K + m_N)\,a^{K^-p}_0\,
  \frac{ie}{2\mu \omega}\,
  \vec{e}^{\;*}(\vec{p},\lambda)\cdot \vec{q}\nonumber\\
  &&\times\, e^{\textstyle\,\pi/2ka_B}\,\Gamma(1 - i/ka_B)\,
  e^{\textstyle\,\pi/2qa_B}\,\Gamma(2 - i/qa_B),
\end{eqnarray}
where $a_B = 1/\alpha \mu$ is the Bohr radius and the
$\Gamma$--functions are defined by \cite{MA72}
\begin{eqnarray}\label{label2.6}
  e^{\textstyle\,\pi z/2}\,\Gamma(1 - i z) &=& 
  \sqrt{\frac{2\pi z}{\displaystyle 1 - e^{\textstyle\, -\,2\pi\,z}}}\,
  \exp\Big\{i\Big[\gamma\,z + \sum^{\infty}_{k = 1}\Big(\frac{z}{k} -
  \arctan\Big(\frac{z}{k}\Big)\Big]\Big\},\nonumber\\
  e^{\textstyle\,\pi z/2}\,\Gamma(2 - i z) &=& (1 - iz)\,
e^{\textstyle\,\pi z/2}\,\Gamma(1 - i z)
\end{eqnarray}
with Euler's constant $\gamma = 0.57721 \ldots$ \cite{MA72}.

For the calculation of the amplitude Eq.(\ref{label2.5}), taking into
account the Coulomb interaction of the $K^-p$ pair in the initial and
final state, we have used the potential model approach. In this
approach we describe strong low--energy interactions by the effective
zero--range potential
\begin{eqnarray}\label{label2.7}
  V(\vec{r}\,) = - \frac{2\pi}{\mu}\,a^{K^-p}_0\,\delta^{(3)}(\vec{r}\,),
\end{eqnarray}
which is equivalent to the effective local strong low--energy
$KKpp$--interaction ${\cal L}_{KKpp}(x) = 4\pi\,(1 + m_K/m_N)\,
a^{K^-p}_0\, \bar{p}(x)p(x)K^{-\dagger}(x)K^-(x)$ in
Eq.(\ref{label2.3}).

Including the Coulomb interaction the term
$i\,\vec{e}^{\;*}(\vec{p},\lambda)\cdot \vec{q}$ in Eq.(\ref{label2.4})
becomes replaced by
\begin{eqnarray}\label{label2.8}
  \hspace{-0.3in}&&i\vec{e}^{\;*}(\vec{p},\lambda)\cdot \vec{q} \to 
  \vec{e}^{\;*}(\vec{p},\lambda)\cdot \int d^3x\,
  \delta^{(3)}(\vec{r}\,)\bigtriangledown
  \psi^{\,C}_{K^-p}(\vec{q},\vec{r}\,) = i\vec{e}^{\;*}(\vec{p},\lambda
  )\cdot\vec{q}\,e^{\textstyle\,\pi/2qa_B}
  \Gamma(2 - i/qa_B),\nonumber\\
  \hspace{-0.3in}&&
\end{eqnarray} 
where $\psi^{\,C}_{K^-p}(\vec{q},\vec{r}\,)$ is the exact
non--relativistic Coulomb wave function of the relative motion of the
$K^-p$ pair in the incoming scattering state with relative momentum
$\vec{q}$. It is given by \cite{LL65} 
\begin{eqnarray}\label{label2.9}
  \psi^C_{K^-p}(\vec{q},\vec{r}\,) =  e^{\textstyle\,\pi/2qa_B}\,
  \Gamma(1 - i/qa_B)\,
  e^{\textstyle\,i\,\vec{q}\cdot \vec{r}}\,F(i/qa_B,1, iqr -
  i\,\vec{q}\cdot \vec{r}\,).
\end{eqnarray} 
Here $F(i/qa_B,1, iqr - i\,\vec{q}\cdot \vec{r}\,)$ is the confluent
function \cite{LL65,MA72}.  

The factor $ e^{\textstyle\,\pi/2ka_B}\,\Gamma(1 - i/ka_B)$ in
Eq.(\ref{label2.5}) is the rest of the asymptotic of the
non--relativistic Coulomb Green function
$G^{\,C}_{K^-p}(\vec{r},0;k^2)$ \cite{JM33}
\begin{eqnarray}\label{label2.10}
  G^{\,C}_{K^-p}(\vec{r}, 0; k^2) = -\,\frac{1}{4\pi r}\,
  \Gamma(1 - i/ka_B)\,W_{i/ka_B,1/2}(-\,2ik r)
\end{eqnarray} 
of the relative motion of the $K^-p$ pair at $r \to \infty$, where
$W_{i/ka_B,1/2}(-\,2ik r)$ is the Whittaker function \cite{MA72},
describing the outgoing spherical wave distorted by the Coulomb
interaction \cite{JM33}.

For the amplitude of the transition $(K^-p)_{np} \to (K^-p)_{1s} +
\gamma$ we get
\begin{eqnarray}\label{label2.11}
  \hspace{-0.3in} &&M((K^-p)_{np} \to (K^-p)_{1s}\, \gamma) = 
8\pi\,(m_K + m_N)\,a^{K^-p}_0\, 
  \frac{ i e}{4\mu^2 \omega}\,\vec{e}^{\,*}(\vec{p},\lambda)
  \cdot \nonumber\\
  \hspace{-0.3in} &&\int \frac{d^3k}{(2\pi)^3}\,
  \Phi^*_{100}(\vec{k})\,e^{\textstyle\,\pi/2ka_B}\,\Gamma(1 -
  i/ka_B)\int \frac{d^3q}{(2\pi)^3}\, 
  e^{\textstyle\,\pi/2qa_B}\,\Gamma(2 - i/qa_B)\,\vec{q}\,
  \Phi_{n1m}(\vec{q}\,).\nonumber\\
  \hspace{-0.3in} &&
\end{eqnarray}
The main contributions to the integrals over $\vec{k}$ and $\vec{q}$
come from the regions $k \ge 1/a_B$ and $q > 1/n a_B$ \cite{TE88}.
These momenta are of order of $O(\alpha)$.  Therefore, the
contribution of the photon momentum can be dropped, since it is of
order of $O(\alpha^2)$.

For the integration over $\vec{q}$ we define the vector
$\vec{q}$ in the spherical basis \cite{IV4}
\begin{eqnarray}\label{label2.12}
  \vec{q} = q\,\sqrt{\frac{4\pi}{3}}\sum_{M = 0,\pm 1}Y^*_{1M}
(\vartheta_{\vec{q}},\varphi_{\vec{q}}\,)\,\vec{e}_M,
\end{eqnarray}
where $\vec{e}_{\pm 1} = \mp (\vec{e}_x \pm i\vec{e}_y)/\sqrt{2}$ and
$\vec{e}_0 = \vec{e}_z$ are spherical unit vectors expended into
Cartesian unit vectors $\vec{e}_x$, $\vec{e}_y$ and $\vec{e}_z$. Using
Eq.(\ref{label2.12}) for the integral over $\vec{q}$ we get
\cite{IV2}
\begin{eqnarray}\label{label2.13}
  \hspace{-0.3in} &&\int \frac{d^3q}{(2\pi)^3} e^{\textstyle\,\pi/2qa_B}
  \Gamma(2 - i/qa_B)\vec{q}\,\Phi_{n1m}(\vec{q}\,) 
  = - i 
  \frac{\vec{e}_m}{\sqrt{3}}\int \frac{d^3q}{(2\pi)^3}
  e^{\textstyle\,\pi/2qa_B}\Gamma(2 - i/qa_B)\,q
  \,\Phi_{np}(q).\nonumber\\
 \hspace{-0.3in} &&
\end{eqnarray}
For the integration over $k$ and $q$ we use the wave functions 
$\Phi_{100}(\vec{k}\,) = \Phi_{1s}(k)$
 and $\Phi_{np}(q)$ given by \cite{BS57}
\begin{eqnarray}\label{label2.14}
  \hspace{-0.3in}  \Phi_{1s}(k) =\frac{8\sqrt{\pi a^3_B}}{(1 + k^2 a^2_B)^2}\;,\;
  \Phi_{np}(q)\sqrt{\frac{\pi n^3 a^3_B}{n^2 - 1}}\,
  \frac{32\,n q a_B}{(1 + n^2q^2a^2_B)^3}\,C^2_{n - 2}
  \Big(\frac{n^2 q^2 a^2_B - 1}{n^2 q^2 a^2_B + 1}\Big),
\end{eqnarray}
where $C^2_{n - 2}(z)$ is the Gegenbauer polynomial \cite{BS57,MA72}.

Using \cite{MTH} and taking into account the results obtained in
\cite{IV2} we make the integration over momenta $k$ and $q$:
\begin{eqnarray}\label{label2.15}
  \hspace{-0.3in} \int \frac{d^3k}{(2\pi)^3}\,\Phi^*_{1s}(k)\,
  e^{\textstyle\,\pi/2ka_B}\,
  \Gamma(1 - i/ka_B) &\simeq&\xi_{1s}\,
  \sqrt{\frac{1}{\pi a^3_B}}\,e^{\textstyle\,i\,\varphi_{1s}},\nonumber\\
  \hspace{-0.3in} \int \frac{d^3q}{(2\pi)^3}\,e^{\textstyle\,\pi/2qa_B}\,
  \Gamma(2 - i/qa_B)\,q\,\Phi_{np}(q) &\simeq&\xi_{np}\,
  \sqrt{\frac{1}{\pi a^5_B}\,\frac{n^2 - 1}{ n^5}}\,
  e^{\textstyle\,i\,\varphi_{np}},
\end{eqnarray}
where $\xi_{1s} = 1.91$ and $\xi_{2p} = 3.52, \xi_{3p} = 2.22,
\xi_{4p} = 2.85, \ldots$ and $\varphi_{1s}$ and $\varphi_{np}$ are
real phases, which do not contribute to the rate of the transition
$(K^-p)_{np} \to (K^-p)_{1s} + \gamma$. 

The amplitude of the transition $(K^-p)_{np} \to (K^-p)_{1s} +
\gamma$ reads
\begin{eqnarray}\label{label2.16}
  &&M((K^-p)_{np} \to (K^-p)_{1s}\, \gamma) = 8\pi\,(m_K + m_N)\nonumber\\
  &&\times\,i\,a^{K^-p}_0\,
  \frac{\mu^2\,}{\omega}\,\sqrt{\frac{\alpha^9}{4\pi}}\,
  \vec{e}^{\,*}(\vec{p},\lambda)\cdot 
  \frac{\vec{e}_m}{\sqrt{3}}\,\sqrt{\frac{n^2 - 1}{n^5}}\,\xi_{1s}\xi_{np}\,
  e^{\textstyle\,i(\varphi_{1s} + \varphi_{np})}.
\end{eqnarray}
The rate of the transition $(K^-p)_{np} \to (K^-p)_{1s} + \gamma$ is
equal to
\begin{eqnarray}\label{label2.17}
  \Gamma((K^-p)_{np} \to (K^-p)_{1s}\, \gamma) = \frac{8}{n^3}\,
\frac{\xi^2_{np}}{\xi^2_{2p}}\,\Gamma((K^-p)_{2p} \to (K^-p)_{1s}\, \gamma),
\end{eqnarray}
where the rate $\Gamma((K^-p)_{2p} \to (K^-p)_{1s}\, \gamma)$ is defined by
\begin{eqnarray}\label{label2.18}
  \Gamma((K^-p)_{2p} \to (K^-p)_{1s}\, \gamma) = \frac{\xi^2_{1s}\xi^2_{2p}}{9}\,
\alpha^7\mu^3\,|a^{K^-p}_0|^2.
\end{eqnarray}
For the subsequent analysis of the rate of the transition $(K^-p)_{np}
\to (K^-p)_{1s} + \gamma$ it is convenient to represent
$|a^{K^-p}_0|^2$ in terms of the energy level displacement of the
ground state of kaonic hydrogen
\begin{eqnarray}\label{label2.19}
  |a^{K^-p}_0|^2 = \frac{1}{4\alpha^6 \mu^4}\,\Big(\epsilon^2_{1s}
 + \frac{1}{4}\,\Gamma^2_{1s}\Big).
\end{eqnarray}
This is the model--independent  DGBTT (Deser, Goldberger, Baumann,
Thirring \cite{DT54} and Trueman \cite{TT61}) formula. Substituting
Eq.(\ref{label2.19}) into Eq.(\ref{label2.18}) we get
\begin{eqnarray}\label{label2.20}
  \Gamma((K^-p)_{2p} \to (K^-p)_{1s}\, \gamma) &=& 
  \frac{\xi^2_{1s}\xi^2_{2p}}{36}\,
  \frac{\alpha}{\mu}\,\Big(\epsilon^2_{1s} + \frac{1}{4}\,
  \Gamma^2_{1s}\Big). 
\end{eqnarray}
Using the numerical values of the parameters $\xi_{1s}$ and $\xi_{2p}$
we obtain
\begin{eqnarray}\label{label2.21}
  \Gamma((K^-p)_{2p} \to (K^-p)_{1s}\, \gamma) =4.3\times 10^4
  \,\Big(\epsilon^2_{1s} + \frac{1}{4}\,\Gamma^2_{1s}\Big)\,{\rm sec^{-1}},
\end{eqnarray}
where $\epsilon_{1s}$ and $\Gamma_{1s}$ are measured in ${\rm eV}$. 

The recent theoretical value of the energy level displacement of the
ground state of kaonic hydrogen reads \cite{IV3}
\begin{eqnarray}\label{label2.22}
  -\,\epsilon_{1s} + i\,\frac{\Gamma_{1s}}{2} = (- 203 \pm 15) +
 i\,(113 \pm 14)\,{\rm eV}.
\end{eqnarray}
Inserting Eq.(\ref{label2.22}) into Eq.(\ref{label2.21}) for the rate
of the transition $\Gamma((K^-p)_{2p} \to (K^-p)_{1s}\, \gamma)$ we
get
\begin{eqnarray}\label{label2.23}
  \Gamma((K^-p)_{2p} \to (K^-p)_{1s}\, \gamma) = 
      (2.3\pm 0.3)\times 10^9\,{\rm sec^{-1}}
\end{eqnarray}
According to Eq.(\ref{label2.17}), the rate of the transition
$(K^-p)_{3p} \to (K^-p)_{1s} + \gamma$ is equal to
\begin{eqnarray}\label{label2.24}
  \Gamma((K^-p)_{3p} \to (K^-p)_{1s}\, \gamma) = 
      (2.7\pm 0.4)\times 10^8\,{\rm sec^{-1}}.
\end{eqnarray}
These rates should be compared with the rates of the pure electric
dipole transitions $2p \to 1s + \gamma$ and $3p \to 1s + \gamma$ at
the neglect of strong interactions.

Using the results obtained by Bethe and Salpeter \cite{BS57} and
adjusting them to kaonic hydrogen we get $\Gamma_{2p \to 1s} =
4.0\times 10^{11}\,{\rm sec^{-1}}$ and $\Gamma_{3p \to 1s} = 1.0\times
10^{11}\,{\rm sec^{-1}}$, respectively. Hence, the rates of the
transitions $(K^-p)_{2p} \to (K^-p)_{1s} + \gamma$ and $(K^-p)_{3p}
\to (K^-p)_{1s} + \gamma$, given by Eqs.(\ref{label2.23}) and
(\ref{label2.24}) and induced by strong low--energy interactions, make
up about $0.6\,\%$ and $0.3\,\%$ of the rates of the pure electric
dipole transitions $2p \to 1s + \gamma$ and $3p \to 1s + \gamma$,
respectively.

For the experimental values of the shift and width of the energy level
of the ground state of kaonic hydrogen, measured by Iwasaki {\it et
  al.} (the KEK Collaboration) \cite{KEK} : $(\epsilon_{1s},
\Gamma_{1s}) = (- 323 \pm 64, 407 \pm 230)\,{\rm eV}$, the rates of
the transitions $(K^-p)_{2p} \to (K^-p)_{1s} + \gamma$ and
$(K^-p)_{3p} \to (K^-p)_{1s} + \gamma$, given by Eqs.(\ref{label2.23})
and (\ref{label2.24}), become increased by a factor of three. 

\section{$np \to 1s + \gamma$ transitions in kaonic deuterium}
\setcounter{equation}{0} 

The formula (\ref{label2.20}) can be easily extended to radiative
transitions in kaonic deuterium $(K^-d)_{np} \to (K^-d)_{1s} +
\gamma$, induced by strong low--energy interactions.  This reads
\begin{eqnarray}\label{label3.1}
  \Gamma((K^-d)_{2p} \to (K^-d)_{1s}\,\gamma) =
  \frac{\xi^2_{1s}\xi^2_{2p}}{36}\,
  \frac{\alpha}{\mu}\,\Big(\epsilon^2_{1s} + \frac{1}{4}\,
\Gamma^2_{1s}\Big) = 3.6\times 10^4\,\Big(\epsilon^2_{1s} + \frac{1}{4}\,
\Gamma^2_{1s}\Big)\,{\rm sec^{-1}},
\end{eqnarray}
where $\mu = m_Km_d/(m_K + m_d) = 391\,{\rm MeV}$ is the reduced mass
of the $K^-d$ pair and $m_d = 1876\,{\rm MeV}$ is the deuteron mass.

Recently the energy level displacement of the ground state of kaonic
deuterium has been estimated in \cite{IV4}:
\begin{eqnarray}\label{label3.2}
  -\,\epsilon_{1s} + i\,\frac{\Gamma_{1s}}{2} = 
      (-325 \pm 60) + i\,(315 \pm 50)\,{\rm eV}.
\end{eqnarray}
According to Eq.(\ref{label3.2}), the rates of the radiative
transitions $(K^-d)_{2p} \to (K^-d)_{1s} + \gamma$ and $(K^-d)_{3p}
\to (K^-d)_{1s} + \gamma$ are equal to
\begin{eqnarray}\label{label3.3}
  \Gamma((K^-d)_{2p} \to (K^-d)_{1s}\, \gamma) &=& 
(7.4\pm 1.8)\times 10^9\,{\rm sec^{-1}},\nonumber\\
  \Gamma((K^-d)_{3p} \to (K^-d)_{1s}\, \gamma) &=& 
(8.7 \pm 2.1)\times 10^8\,{\rm sec^{-1}}.
\end{eqnarray}
These rates should be compared with the rates $2p \to 1s + \gamma$ and
$3p \to 1s + \gamma$ of the pure electric dipole transitions. Using
the results obtained by Bethe and Salpeter \cite{BS57} and adjusting
them to kaonic deuterium we get $\Gamma_{2p \to 1s} = 4.8\times
10^{11}\,{\rm sec^{-1}}$ and $\Gamma_{3p \to 1s} = 1.2\times
10^{11}\,{\rm sec^{-1}}$, respectively.

Thus, the rates of the radiative transitions $(K^-d)_{2p} \to
(K^-d)_{1s} + \gamma$ and $(K^-d)_{3p} \to (K^-d)_{1s} + \gamma$,
induced by strong low--energy interactions, make up about $1.5\%$ and
$0.7\,\%$ of the rates of the pure electric dipole transitions $2p \to
1s + \gamma$ and $3p \to 1s + \gamma$, respectively.

For the value of the shift and width of the energy level of the ground
state of kaonic deuterium, predicted by Barrett and Deloff
\cite{RB99}: $ (\epsilon^{(d)}_{1s}, \Gamma^{(d)}_{1s}) = (-\,693,
880)\,{\rm eV}$, the rates of the transitions $(K^-d)_{2p} \to
(K^-d)_{1s} + \gamma$ and $(K^-d)_{3p} \to (K^-d)_{1s} + \gamma$,
given by Eq.(\ref{label3.3}), become increased by more than three
times.

\section{Conclusion}

We have calculated the rates of the radiative transitions $np \to 1s +
\gamma$ for kaonic hydrogen and kaonic deuterium, induced by strong
low--energy interactions and enhanced by the attractive Coulomb
interaction of the $K^-p$ and $K^-d$ pairs in the kaon--proton and
kaon--deuteron {\it Bremsstrahlung}, $K^- + p \to K^- + p + \gamma$
and $K^-+ d \to K^- + d + \gamma$. The neglect of the Coulomb
interaction of the $K^-p$ and $K^-d$ pairs in these reactions
corresponds to $\xi_{1s} = \xi_{np} = 1$ in Eq.(\ref{label2.16}). For
the calculation of the amplitudes of the kaon--proton and
kaon--deuteron {\it Bremsstrahlung} we have used an approach analogous
to that of the EFT based on ChPT by Gasser and Leutwyler \cite{JG83}
and used in \cite{IV4} for the derivation of the Ericson--Weise
formula for the S--wave scattering length of $K^-d$ scattering
\cite{TE88}. It agrees also well with the potential model approach.

We have found that for the $2p$ states of kaonic atoms the
contributions of the rates of the transitions $(K^-p)_{2p} \to
(K^-p)_{1s} + \gamma$ and $(K^-d)_{2p} \to (K^-d)_{1s} + \gamma$,
induced by strong low--energy interactions, relative to the rates of
the pure electric dipole transitions $2p \to 1s + \gamma$, are of
order of one percent. 
 
Precisions of this order $\pm\,0.2\,\%$ and $\pm\,1.0\,\%$ have been
reached in the experiments of the PSI Collaboration \cite{PSI1} for
the measurements of the energy level shift and width of the ground
state of pionic hydrogen, respectively. These precisions are defined
by the accuracy of the measurements of the $X$--ray spectra and yields
in pionic hydrogen.

Measurements of the energy level displacements of the ground states
for kaonic atoms at the same level of precision, would demand to take
into account the rates of the transitions $(K^-p)_{np} \to (K^-p)_{1s}
+ \gamma$ and $(K^-d)_{np} \to (K^-d)_{1s} + \gamma$, induced by
strong low--energy interactions, for the theoretical description of
the experimental data on the $X$--ray spectra and yields in kaonic
atoms.

\section*{Acknowledgement}

We are grateful to Torleif Ericson for numerous and fruitful  discussions.

\end{document}